\documentclass[11pt]{article}
\usepackage{moriond,epsfig}

\def\Journal#1#2#3#4{{#1} {\bf #2}, #3 (#4)}

\def\be{\begin{equation}}
\def\ee{\end{equation}}
\def\bea{\begin{eqnarray}}
\def\eea{\end{eqnarray}}

\begin{document}
\vspace*{4cm}
\title{WEAK GRAVITATIONAL LENSING}

\author{ ALAN HEAVENS }

\address{SUPA\footnote{Scottish Universities Physics Alliance}, Institute for Astronomy, University of Edinburgh,\\ Blackford Hill, Edinburgh EH9 3HJ, U.K.}

\maketitle\abstracts{In this brief review I consider the advances made in weak gravitational lensing over the last 8 years, concentrating on the large scales - cosmic shear.  I outline the theoretical developments, observational status, and the challenges which cosmic shear must overcome to realise its full potential.  Finally I consider the prospects for probing Dark Energy and extra-dimensional gravity theories with future experiments.}

\section{Introduction}

Weak lensing refers to the coherent distortion of images of distant objects, caused by the passage of light through the non-uniform mass distribution in the Universe.  It is a particularly valuable probe because it is blind as to the nature of the mass, and is therefore useful for studying Dark Matter.  Less obviously, it is also sensitive to the properties of Dark Energy, as it depends on both the geometry of the Universe and the growth rate of perturbations: the more non-uniform the Universe, the bigger the distortion, but also for a given redshift of source, the longer the path-length, the greater the distortion.   This dual sensitivity is a great advantage, particularly for studies of gravity theories, as it can lift some degeneracies which otherwise exist if one probes only the geometry of the Universe (for example, by Baryon Acoustic Oscillations or supernova Ia observations).   Another theoretical advantage is the fact that the physics is very simple, being dominated by gravity without complex astrophysics.  The exception to this is the shear-intrinsic shape correlation, which we explore later in this review.  Observationally, weak lensing is challenging, with careful measurement and removal of optical distortions being required, and accurate measurement of galaxy shapes.  Finally, in order to exploit the full power of weak lensing measurements, it is necessary to have distance estimates for individual sources, and the accuracy required places stringent constraints on systematic photometric redshift errors.

It is easy to forget that cosmic shear is a very young subject.  The first measurements were published only in 2000 (see Table 1), and progress in theory, observation and control of systematics is rapid.  Nevertheless, weak lensing is still catching up with other cosmological probes, and is not yet quite at the stage of providing the most stringent constraints on cosmological parameters.  For a recent full review, see Munshi et al.~\cite{MVWH}.

\begin{table}[t]
\caption{Weak lensing: the Bush years.\label{History}}
\vspace{0.4cm}
\begin{center}
\begin{tabular}{|l|l|}
\hline
2000 & First detections~\cite{Bacon2000,Kaiser,Wittman,vWetal}\\
2002+ & Weak-lensing selected cluster catalogues~\cite{Miyazaki,Wittman04}\\
2003+ & Dark matter power spectrum~\cite{Brown03,Heymans04,Hoekstra04,Semboloni06}\\
2004 & Bullet cluster challenge to MOND~\cite{Clowe04}\\
2004+ & 3D potential reconstruction~\cite{Taylor07,Massey07}\\
2005 & Evolution of structure~\cite{Bacon05}\\
2006+ & 3D analyses~\cite{Heavens03,HKT,Kitching08a,Taylor07}\\
2007 & 100 square degree surveys, small errors~\cite{Benjamin07,Fu08}\\
\hline
\end{tabular}
\end{center}
\end{table}

\subsection{Theory}\label{subsec:theory}

Assuming General Relativity, and some weak conditions on the constituents of the Universe, the equation of motion for a photon is given in flat space by $
d^2 {\bf x}/d\eta^2 = -2\nabla\Phi/c^2$,
where $\eta$ is the conformal time $d\eta = dt/R(t)$, $R(t)$ is the cosmic scale factor, and $\Phi$ is the peculiar Newtonian gravitational potential. $\nabla$ here is a comoving transverse gradient operator
$(\partial_x, \partial_y)$.

The distortion of the source (coordinates $\beta_i$) to image ($\theta_j$) is given by the distortion matrix
\[
A_{ij} \equiv \frac{\partial \beta_i}{\partial \theta_j} = \delta_{ij}-\phi_{,ij} = \left( \begin{array}{cc}
1-\kappa & 0 \\
0 & 1-\kappa \end{array} \right) + \left( \begin{array}{cc}
-\gamma_1 & -\gamma_2 \\
-\gamma_2 & \gamma_1 \end{array} \right)
\]
where $\kappa$ is the convergence, which controls the image size (and brightness, since lensing preserves surface brightness), and $\gamma=\gamma_1+i\gamma_2$ is the complex shear, which distorts the shapes.
$\phi$ is the cosmological lensing potential, $\phi({\bf r}) \equiv 2\int_0^r dr' (r-r')
\Phi({\bf r}')/(c^2 r r')$,
and $\phi_{,ij}({\bf r}) \equiv \partial^2 \phi({\bf r})/\partial \theta_i
\partial \theta_j$. The integral is understood to be along a radial line (Born approximation), which is
a very good approximation for weak lensing
~\cite{Bernardeau,Schneider98,VW2002}.

The convergence and shear are therefore related to second derivatives of the potential, which means there should be only E-modes (to a good approximation), and no B-modes, which provides a useful check on systematics. For lensing dominated by a single object along the line-of-sight, the convergence is proportional to the surface density of the lens.  Shear is normally used for cosmological parameter estimation, rather than size or brightness, as it has higher signal-to-noise.  Note that typically $\kappa$ and $\gamma$ are only $\sim 0.01$.

\subsection{Sensitivity to Dark Energy}

The lensing potential evidently probes both the growth rate (via $\Phi$), and the distance-redshift relation $r(z)$, since $r$ is not directly observable.
The radial distance is related to the Hubble expansion parameter $H(t)\equiv R^{-1}dR/dt$ by
$r(z) = c\int_0^z\,dz'/H(z'),$
For an equation of state parameter $w\equiv p/\rho$ for Dark Energy which changes with scale factor, $w(a)$, the Hubble parameter is given by
\begin{equation}
H^2(a) = H_0^2\left[\Omega_m a^{-3}+\Omega_k a^{-2}+\Omega_{DE}
\exp\left(3\int_1^a\,\frac{da'}{a'}\left[1+w(a')\right]\right)\right]
\end{equation}
where $\Omega_m$, $\Omega_{DE}$ and $\Omega_k$ are the present matter, Dark Energy and curvature density parameters.  We see therefore that the Dark Energy sensitivity is via the Hubble parameter, or equivalently through its effect on the expansion history of the Universe.

The Dark Energy also affects the growth rate via the Hubble parameter, since in General Relativity, the fractional overdensity $\delta \equiv \delta \rho/\bar \rho-1$ (where $\bar\rho$ is the mean density) grows to linear order according to
$\ddot\delta+2 H \dot\delta -4\pi G \rho_m \delta = 0,$
where $\rho_m$ is the matter density and we assume the Dark Energy density is not perturbed.

\section{Shape measurement and shear estimation}

Apparent shapes of stars are used to correct for PSF distortions, and very accurate shape measurement is required. The industry standard for measuring shapes is KSB~\cite{KSB}, which defines an ellipticity in terms of the moments of the surface brightness distribution.  For any shape measurement statistic, one needs to know how it is changed under shear.  For one definition~\cite{SeitzSchneider95}, the complex ellipticity $e$ transforms according to
$e = (e_s+g)/(1+g^*e_s)$,
where $e_s$ is the source ellipticity, and $g=\gamma/(1-\kappa)$ is the reduced shear.
If we average over many galaxies,
$\langle e \rangle = g$.  Note that $e$ is dominated by the intrinsic
ellipticity, and many source galaxies are needed to get a robust
measurement of cosmic shear.  Large-area surveys are therefore required for high accuracy, and substantial depth, as the lensing signal drops rapidly for sources at $z<0.5$.   The requirements on accurate shape measurement are quite severe for the error on Dark Energy properties not to be dominated by it.  The ellipticities need to be measured with a systematic error which is rather less than 1\% of the shear signal - i.e. to a systematic accuracy of $<10^{-4}$.  This seems achievable with the latest shape measurement methods such as lensfit~\cite{Miller07,Kitching08b}.

\section{Current observational status}

\subsection{2D and 3D mass reconstruction}

\begin{figure}[hb]
\begin{minipage}[b]{0.45\linewidth} 
\centering \psfig{figure=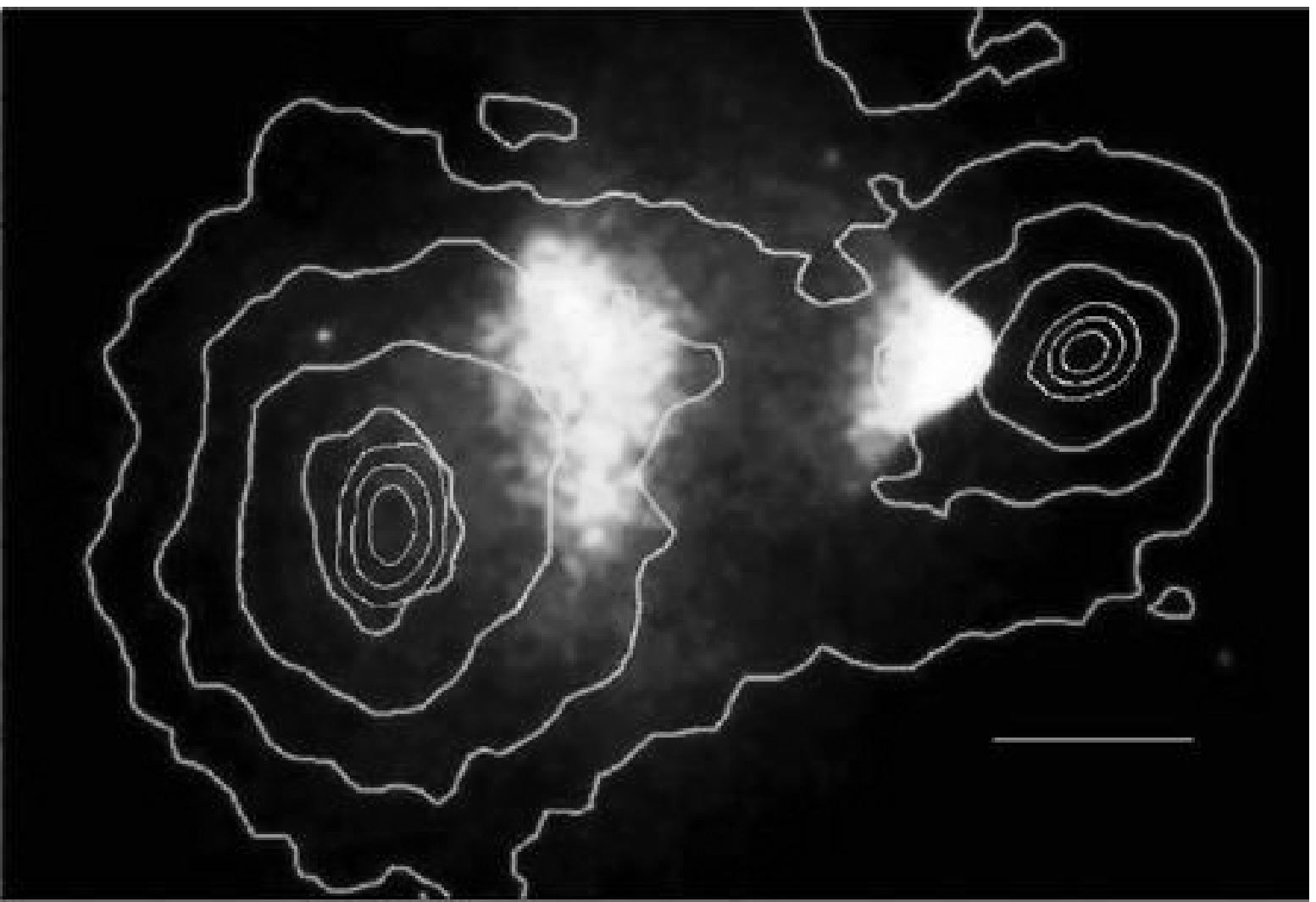,width=6 cm}
\caption{Bullet cluster in X-rays (image) and surface mass density (contours), as measured by Clowe et al (2004) from weak lensing.}
\label{Bullet}
\end{minipage}
\hspace{0.5cm} 
\begin{minipage}[b]{0.45\linewidth}
\centering \psfig{figure=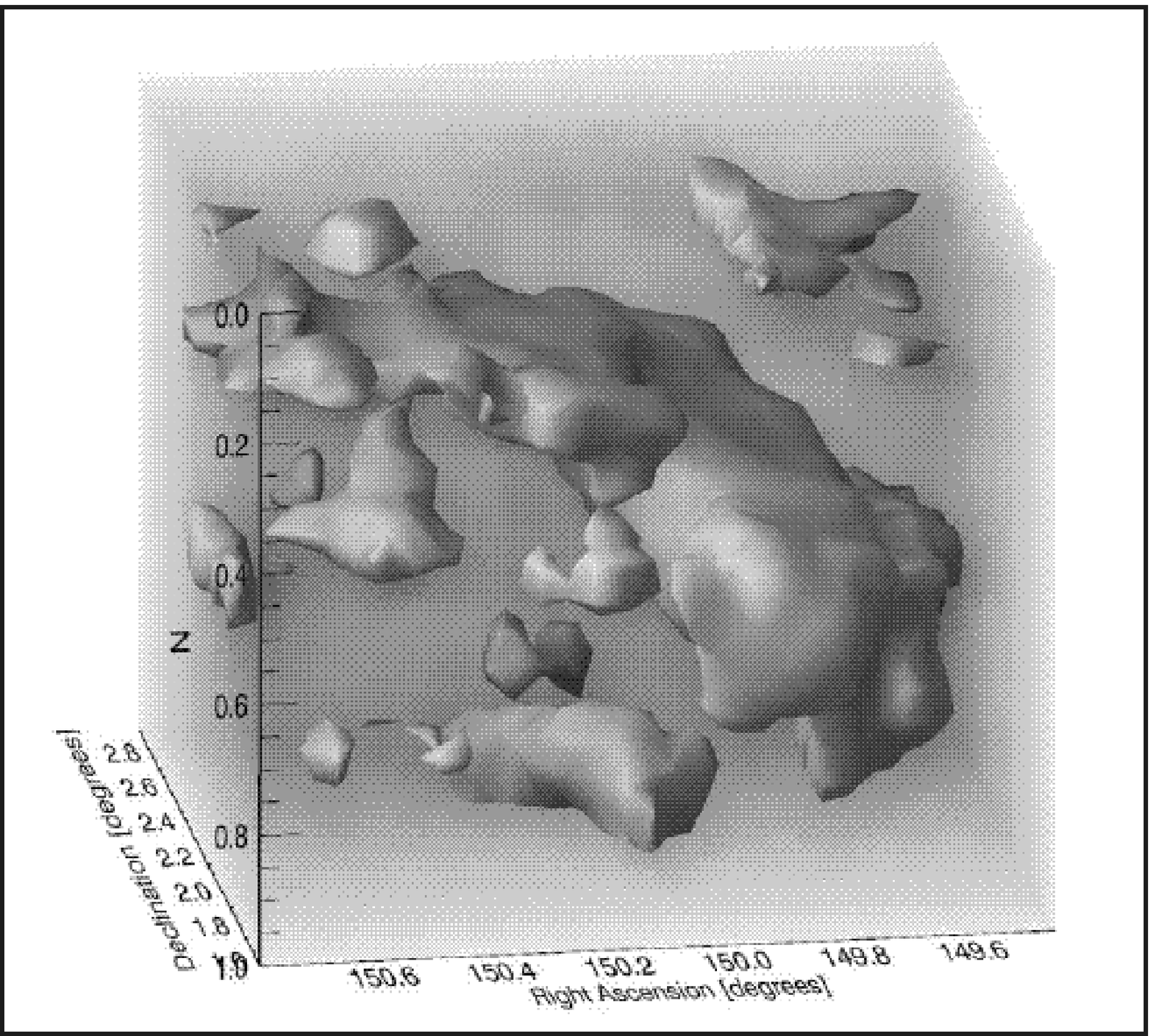,width=6 cm}
\caption{3D reconstruction of matter density from the COSMOS ACS data (Massey et al 2007).}
\label{COSMOS3D}
\end{minipage}
\end{figure}

It has been known for many years that weak lensing data can be used to measure the surface mass density, and this has perhaps been used to greatest effect~\cite{Clowe04} with the Bullet cluster (Fig. \ref{Bullet}).  This image, showing two clusters after a recent collision, demonstrates that the main baryonic material, the x-ray emitting gas, is displaced from the concentrations of convergence (contours).  This picture is consistent with the standard model of cosmology with collisionless Dark Matter, and presents difficulties for the MOND/TeVeS theories, but note that the convergence is not proportional to the surface density in such theories.

Remarkably, the weak shear data from galaxies with distance information can be inverted~\cite{Taylor01} to yield the 3D gravitational potential and hence the matter density.  This method was first applied to COMBO-17 data~\cite{Taylor04}, and recently to COSMOS HST data~\cite{Massey07} - see Fig.\ref{COSMOS3D}.

\subsection{Cosmological parameter estimation}

\begin{figure}[hb]
\begin{minipage}[b]{0.45\linewidth} 
\centering \psfig{file=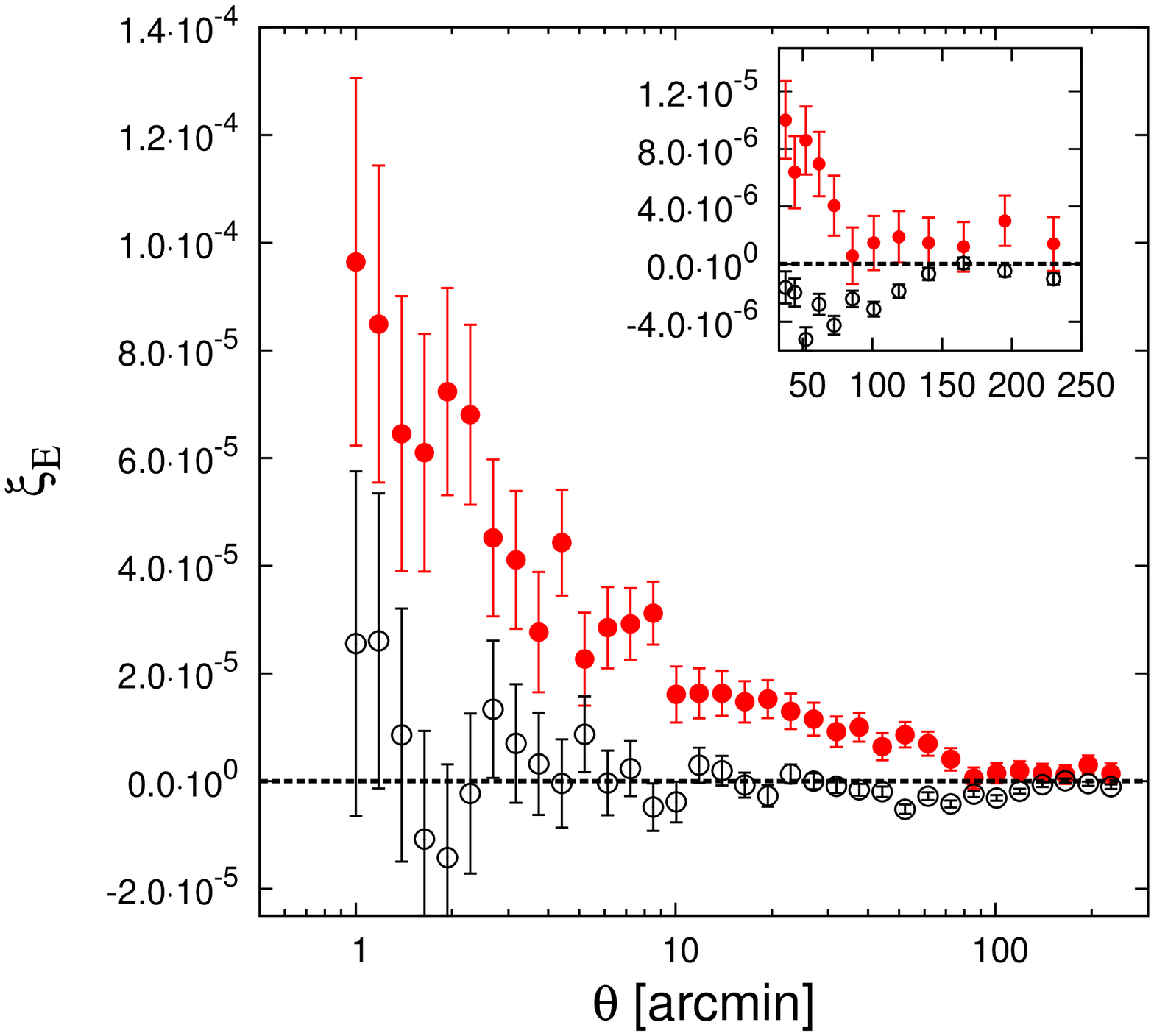,width=5 cm,height=5 cm, angle=270}
\caption{Shear correlation function from the CFHTLS wide survey (Fu et al 2008).}
\label{CFHTLS}
\end{minipage}
\hspace{0.5cm} 
\begin{minipage}[b]{0.45\linewidth}
\centering \psfig{file=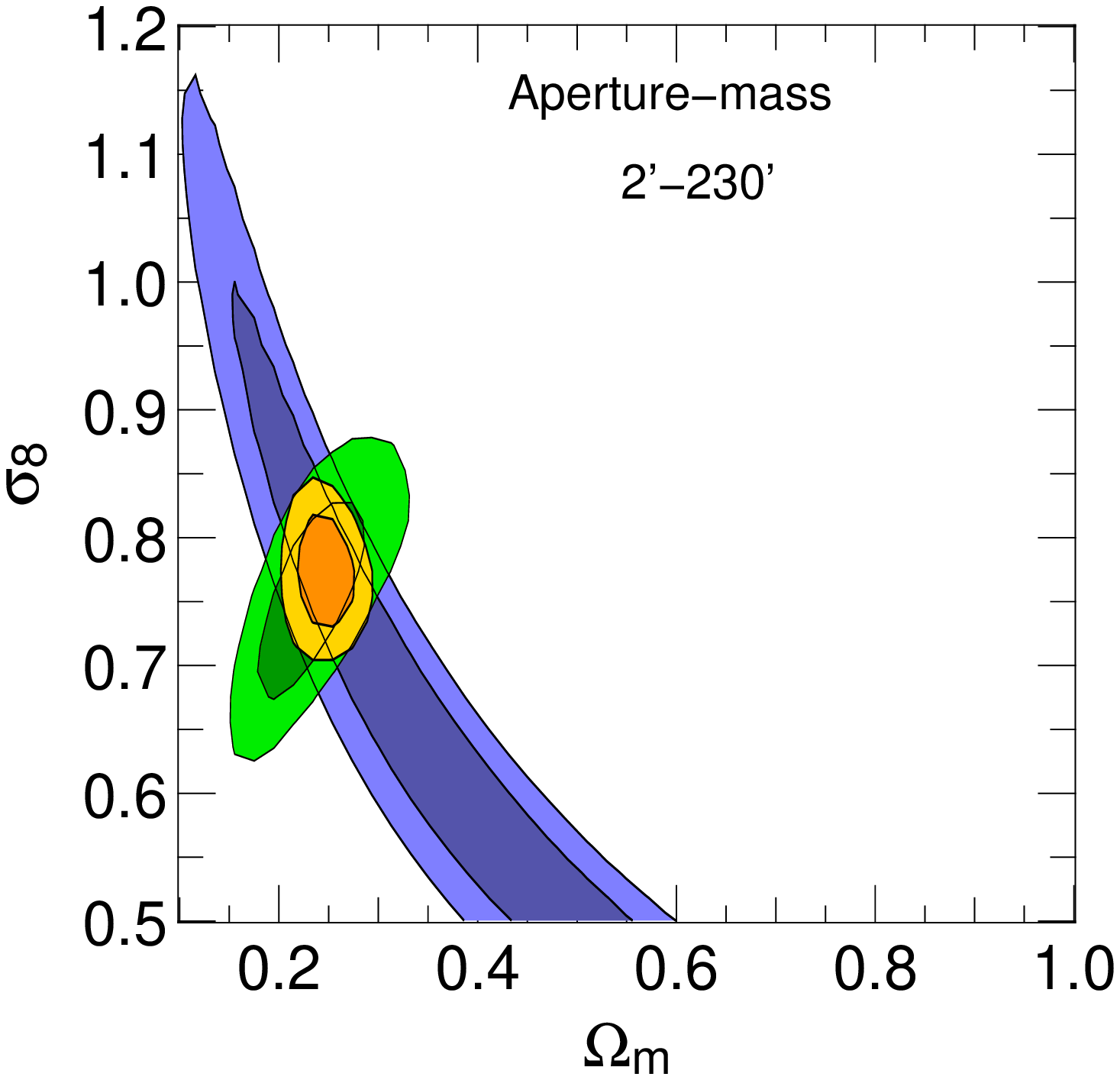,width=5 cm,height=5 cm}
\caption{Cosmological parameters from CFHTLS (Fu et al 2008) and WMAP (Spergel et al. 2007).}
\label{CFHTLSWMAP}
\end{minipage}
\end{figure}

In many respects the Canada-France-Hawaii-Telescope legacy survey (CFHTLS) represents the state-of-the-art as far as ground-based weak lensing surveys are concerned.  The latest results~\cite{Fu08} from 57 square degrees are represented by Fig. \ref{CFHTLS}, which shows the angular shear correlation function, including an estimate of the B-modes (open), which are consistent with zero except at the plate scale.  The constraints on $\Omega_m$ and the amplitude of matter fluctuations, $\sigma_8$, are shown in Fig.\ref{CFHTLSWMAP}, showing the normal banana degeneracy expected from a 2D analysis, and compared with WMAP 3-year data.  In combination, they yield $\Omega_m=0.25\pm 0.02$ and $\sigma_8=0.77\pm 0.03$.  Previous tensions with WMAP's $\sigma_8$ have disappeared with better determination of the redshift distribution of the lensed sources, which were previously estimated using the Hubble Deep Field, which has large sample variance~\cite{Benjamin07}.

\section{More Systematics: Intrinsic alignments and photometric redshifts}

The main signature of weak lensing is a small alignment of the
images, at the level of a correlation of ellipticities of $\sim
10^{-4}$.  Physical alignment of nearby galaxies may mimic this, and
was first investigated theoretically~\cite{HRH,CM,Crittenden01,CKB,Jing}, and found to be non-negligible, and observationally~\cite{Brown}.  However, with
photometric redshifts, one can remove galaxies
which may be physically close from pair statistics~\cite{HH,KS}, as was done in~\cite{Heymans05}. Thus one essentially completely removes a systematic error in
favour of a slightly increased statistical error, so this systematic is not a concern.
\begin{figure}[hb]
\begin{minipage}[b]{0.45\linewidth} 
\centering \psfig{file=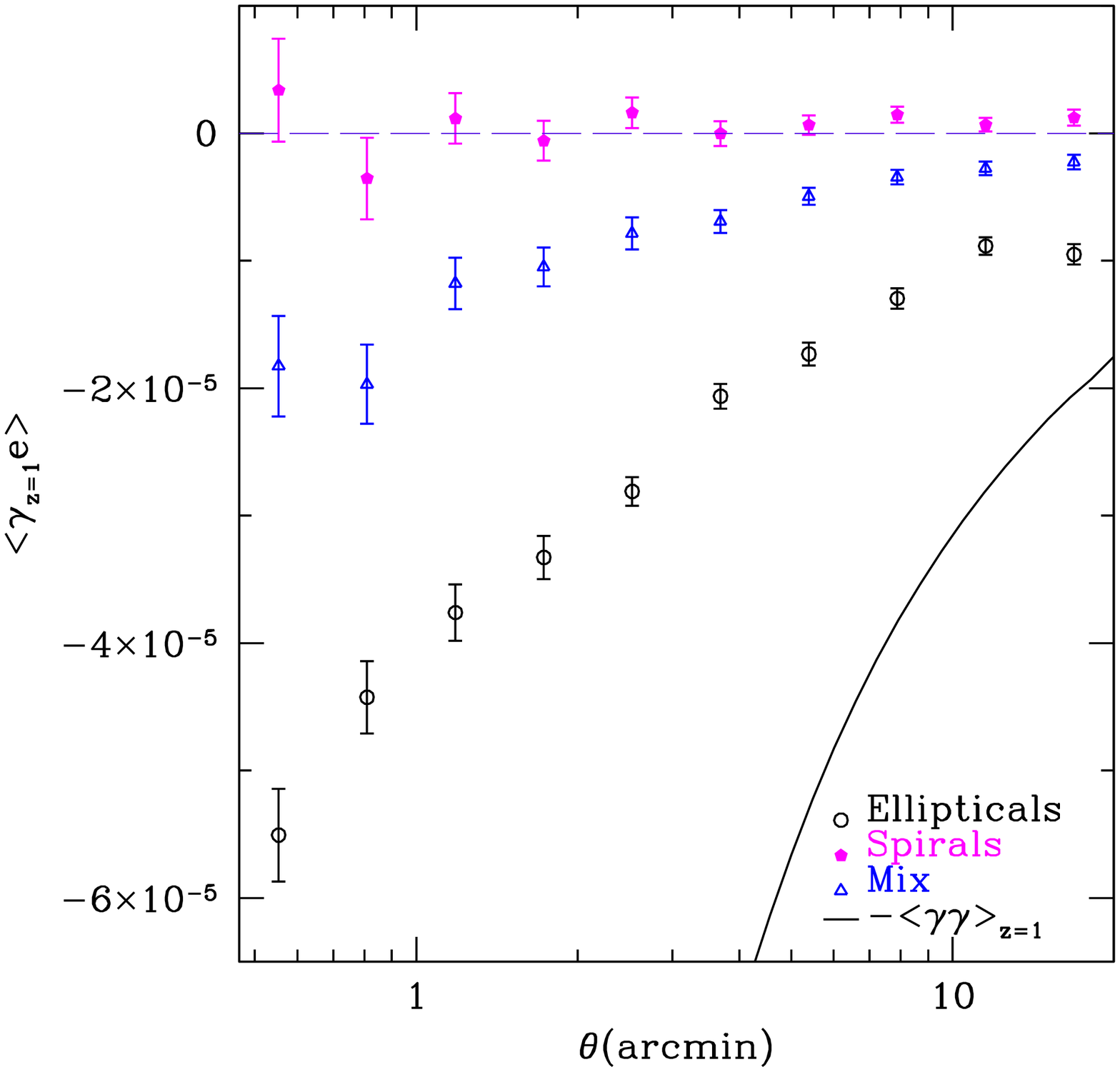,width=2.5in,height=2in}
\caption{N-body shear-intrinsic alignment correlation. Predictions depend on how galaxies are placed in halos: thin disk with angular momentum aligned (top); ellipticity of galaxy same as halo (bottom); mixture (middle).  The solid curve is minus the lensing signal (Heymans et al 2006)}
\label{GI}
\end{minipage}
\hspace{0.5cm} 
\begin{minipage}[b]{0.45\linewidth}
\centering \psfig{file=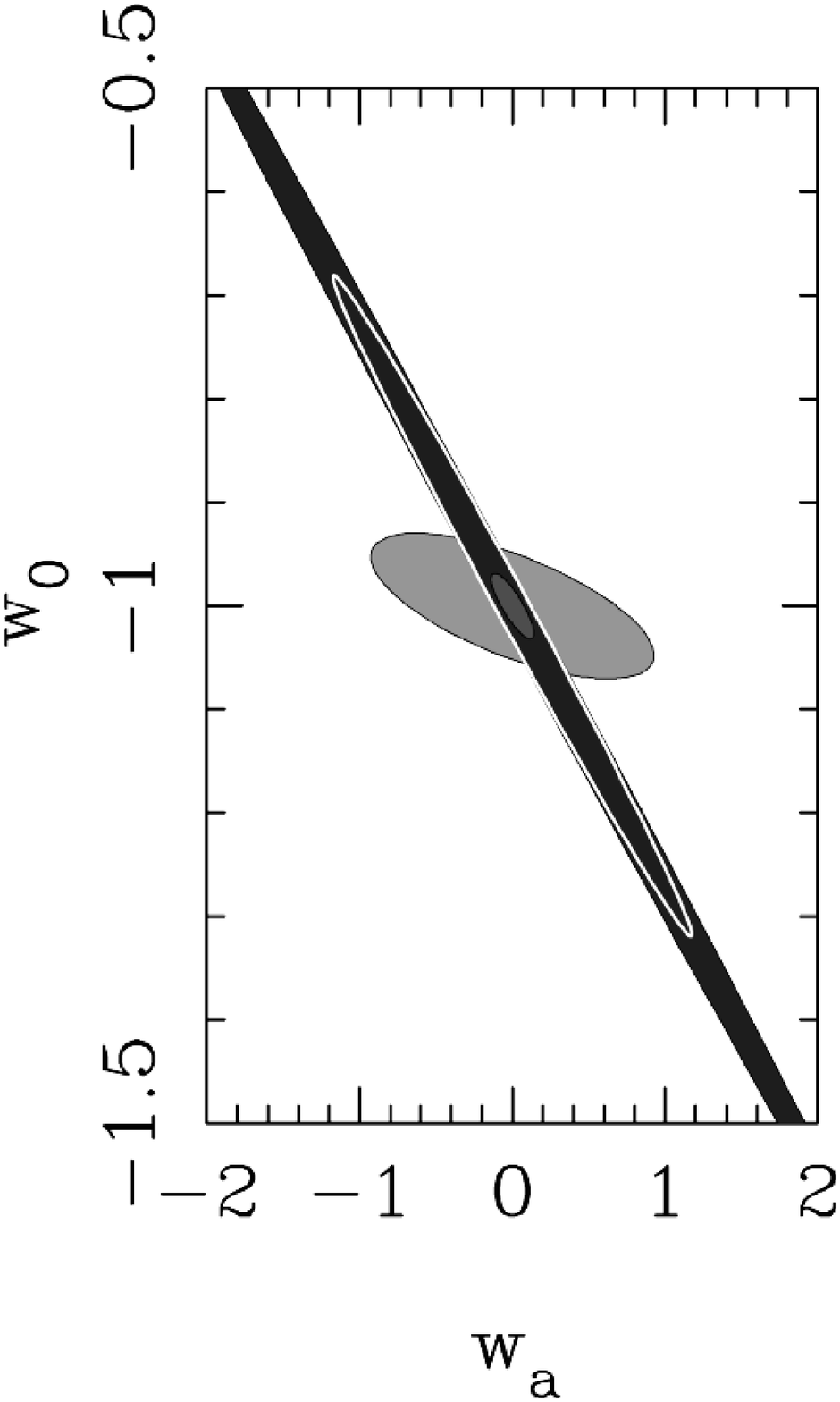,width=2.0in,height=2.8in}
\caption{Expected constraints from Planck (darkest), EUCLID weak lensing (lightest), combined (smallest). From Kitching et al. (2008a).}
\label{Fisher}
\end{minipage}
\end{figure}
More problematic is a subtle correlation between background shear and foreground ellipticity.  This was first pointed out by Hirata \& Seljak~\cite{HS}, and arises if the foreground galaxy is correlated with the local tidal field.  This field contributes to the background shear, and this effect has been seen in simulations~\cite{Heymans06} (Fig.\ref{GI}) and inferred from SDSS observations~\cite{Mandelbaum06,Hirata07}.  It is much more difficult to deal with, as it cannot easily be removed, but it should have a different redshift dependence from lensing, and techniques to deal with it are beginning to emerge~\cite{BridleKing,JoachimiSchneider08}.

We have emphasised the need for photometric redshifts, in order to improve the statistical power, and also to help remove systematics~\cite{BridleKing}.  A major source of error may occur if the photometric redshifts are systematically in error.  This puts severe constraints on the calibration of photometric redshifts, requiring $\sim 10^5$ spectroscopic redshifts.  This is not a fundamental limitation, but an expensive one to fix.  Uncertainty in the highly nonlinear matter power spectrum also limits the range of scales which can be probed.

\section{Future surveys}

A selection of future surveys is presented in Table 2, to which could be added LSST and SNAP.  Given that the largest optical weak lensing surveys are now $\sim 100$ square degrees, the increase by two orders of magnitude is impressive (albeit at shallower depth).  The expected errors on the Dark Energy equation of state parameters (assuming $w(a)=w_0+(1-a)w_a$~\cite{ChevallierPolarski01}) are very small - around a percent or so~\cite{Kitching08a} for $w$ at $z \sim 0.4$, see Fig. \ref{Fisher}.  This high accuracy is possible through treating the shear field in 3D, using photometric redshifts~\cite{Hu99,Heavens03}.

\begin{table}[t]
\caption{Selected future experiments, with areal coverage, depth and number of galaxies with measurable shapes and photo-zs.\label{Future}}
\vspace{0.4cm}
\begin{center}
\begin{tabular}{|l|l|l|l|l|}
\hline
& Area/sq. deg. & Median $z$ & Number density/arcmin$^{-2}$ & Start date\\
\hline
Pan-STARRS 1 & 20000 & $\sim 0.6$ & $>$5 & 2008 \\
KIDS & 1700 & $\sim 0.6$ & $>$5 & 2008\\
DES & 5000 & $\sim 0.7$ & $\sim 10$ & 2010\\
Hypersuprimecam & 2000 & $>$1 & 20-30 & 2013\\
EUCLID (DUNE) & 20000 & $\sim 0.9$ & 40 & 2017\\
\hline
\end{tabular}
\end{center}
\end{table}

An exciting possibility of some of these experiments is that General Relativity could be tested, since modified gravity models, such as those inspired by braneworlds, generically predict a different growth rate from GR, so could in principle be distinguished from GR by weak lensing observations.  Using the convenient Minimal Modified Gravity parametrization~\cite{Linder05},
it has been shown that Pan-STARRS 1 might make a marginal detection of DGP, but EUCLID (formerly named DUNE) should distinguish DGP from GR at high significance~\cite{HKV,Amendola08}.

\end{document}